\documentclass[11pt]{article}
\usepackage{hyperref}

\usepackage{graphicx}
\usepackage{amsmath,amsfonts}
\usepackage{amssymb}

\usepackage{bm}

\def\pr{{\partial}}
\def\eps{{\epsilon}}
\def\bx{{\bf x}}
\def\bv{{\bf v}}
\def\bV{{\bf V}}

\def\ba{{\mathbf{a}}}
\def\bbb{{\mathbf{b}}}
\def\bc{{\mathbf{c}}}
\def\bxi{{\boldsymbol{\xi}}}
\def\bom{{\boldsymbol{\omega}}}

\begin{document}

\title{\textbf{An asymptotic model in acoustics: acoustic drift equations}}

\author{Vladimir A. Vladimirov\footnote{vladimir.vladimirov@york.ac.uk, http://maths.york.ac.uk/www/vv500} 
~and Konstantin Ilin\footnote{konstantin.ilin@york.ac.uk, http://maths.york.ac.uk/www/ki502}\\
{\small Department of Mathematics, University of York, Heslington, York, YO10 5DD, U.K.}}

\maketitle


\begin{abstract}
A rigorous asymptotic procedure with the Mach number as a small parameter is used to derive the equations of mean flows which coexist and are affected by the background acoustic waves in the limit of very high Reynolds number.
\end{abstract}

\maketitle

\section{\label{sec1}Introduction}

In this note we use a rigorous asymptotic procedure to derive new equations for mean flows that coexist and are affected by
a background acoustic field in the limit of very high Reynolds numbers. We will refer to these equations as the {\em acoustic drift equations} because
they contain the drift velocity of fluid particles in the oscillatory acoustic field. Our theory can be viewed as an extension of previous studies on
acoustic streaming (see, e.g., Refs. 1--7) 
to the case of high Reynolds numbers.

The basic parameters of the acoustic streaming flows are the Mach number $\eps$ and the streaming Reynolds number $R_s$ \cite{Lighthill, Riley2001},
defined by
\[
\eps = \frac{V^{*}}{c^{*}_0}, \quad R_s =\frac{V^{*2}\rho^{*}_0}{\omega^{*}\eta},
\]
where $\omega^{*}$ is the frequency of the sound wave,
$V^{*}$ is the amplitude of the velocity oscillations, $\eta$ is the dynamic viscosity, $\rho_0^*$
and $c^{*}_0$ are the density and the velocity of sound in the undisturbed fluid.

The two most popular examples of acoustic streaming are (i) steady flows produced by intense ultrasound beams (`quartz wind') and (ii) steady flows driven by
viscous boundary layers near a rigid boundary. In the first example, it is essential that the sound waves are attenuated due to
viscous dissipation (or due to some other physical mechanism) \cite{Eckart, Westervelt, Nyborg}. The resulting steady flow is described by the Stokes equations 
with an 'external body force' term that appears because of the attenuated sound waves. The theory is restricted to the case of small streaming Reynolds numbers
($R_s\ll 1$).
An interesting set of equations governing unsteady (but slowly varying) mean flows is presented in Chapter 8 of the book by
Rudenko and Soluyan \cite{Rudenko}. The equations are the incompressible Navier-Stokes equations with an external body force that
appears due to the presence of background sound waves. These equations had been used to solve a number of concrete acoustic streaming problems
(see, e.g., Refs. ~\cite{GusevRudenko,Kamakura}) and results seem to be in agreement with observations.
However, if viscous effects are small, this model reduces to the incompressible
Euler equations with a potential external force which can be included into the pressure term. As a result, the mean flow is unaffected
by the sound waves.

In the second example, there is no 'external body force' and the flow is induced by the boundary layer at a rigid boundary. Here there
are theories which can treat acoustic streaming flows at moderate and even high streaming Reynolds numbers \cite{Lighthill, Riley2001}. Acoustic steaming driven by boundary layers will
not be considered in the present note.

The aim of this note is to derive equations that govern the evolution of the mean flow at high
Reynolds numbers, $R_s\gg 1$. To be more precise, we consider flows for which, in the limit of small $\eps$, $R_s\sim \eps^{-\lambda}$
with $\lambda\geq 1$. This means that the viscosity does not affect both the sound waves and the mean flow which are effectively inviscid.
For compressible flows, this limit had not been treated previously.

Our asymptotic procedure is based on the method of multiple scales \cite{Nayfeh} (for an example of using the method in acoustics
see Ref. ~\cite{Nayfeh2}) and is similar to the approach employed
in Refs. ~\cite{Vladimirov2012} and ~\cite{IM2012}. The procedure leads to asymptotic equations describing the mean flow that is quadratic
in the amplitude of the background sound waves. These equations, which we call the acoustic drift equations,
have the form of the incompressible Euler equations with an additional term in the momentum equation
that contains the drift velocity of fluid particles. They give a valid approximation for solutions of the governing equations
everywhere away from rigid walls.
Interestingly, our asymptotic equations are similar to the equations that
had been derived by Craik and Leibovich for incompressible flows and used to explain the Langmuir circulations in the ocean \cite{Craik}.

Our asymptotic procedure can be used to compute as many successive approximations to the solution as necessary. It produces
not only the equations of the mean flow but also the successive approximations for the mean pressure which is needed to find the
radiation force exerted on a rigid particle by the sound waves. Also, an example considered  in Section IV shows that the asymptotic equations derived
in the paper can indeed describe non-trivial mean flows which appear due to the presence of sound waves.

\section{\label{sec2}Formulation of the problem}

Let $\bx^{*}=(x^{*},y^{*},z^{*})$ be Cartesian coordinates in space. We assume heat conduction can be ignored and
consider three-dimensional viscous compressible isentropic flows.
The Navier-Stokes equations
for isentropic flows can be written as \cite{Landau}
\begin{eqnarray}
&&\bv^{*}_{t^{*}}+ (\bv^{*}\cdot\nabla^{*})\bv^{*}=   - \nabla^{*} h^{*} +({\eta}/{\rho}) \,  \hat{K}\bv^{*},  \nonumber \\
&&\rho^{*}_{t^{*}}+ \nabla^{*}\cdot (\rho^{*} \bv^{*}) = 0, \quad
\hat{K}\bv^{*}\equiv \nabla^{*2}\bv +b \, \nabla^{*} (\nabla^{*}\cdot\bv^{*}). \nonumber
\end{eqnarray}
Here $\bv^{*}$ is the velocity of the fluid, $t^{*}$ is time,
$\rho^{*}$ is the density and $h^{*}(\rho^{*})$ is the enthalpy per unit mass of the fluid, $b=({\zeta}/{\eta})+(1/3)$, $\eta$ and $\zeta$
are the shear and bulk viscosities of the fluid. Note that
the velocity of sound $c^{*}$ can be written as
\[
c^{*2}(\rho^{*})=d p^{*}/d \rho^{*}=\rho^{*} \, d h^{*}/d \rho^{*}.
\]
Let $\rho^*_0$
be the undisturbed density, $\omega^{*}$ the angular frequency of a sound wave, $L^{*}$ the wavelength divided by $2\pi$ and $V^{*}$ the amplitude
of the velocity oscillations in the wave. We employ these to define the
non-dimensional quantities $\rho$, $h$, $\bv$, $\bx$ and $\tau$:
\begin{eqnarray}
&&\eps \rho=\frac{\rho^* -\rho^*_0}{\rho^*_0}, \quad
\eps h(1+\eps \rho)=\frac{h^*(\rho^{*})-h^*(\rho^{*}_{0})}{c_0^{*2}}, \nonumber \\
&&\bv^{*}=V^{*}\bv, \quad \bx^{*}=L^{*}\bx, \quad \omega^{*} t^{*}= \tau. \nonumber
\end{eqnarray}
With these variables,
the Navier-Stokes equations become
\begin{eqnarray}
&&\bv_{\tau}+ \eps (\bv\cdot\nabla)\bv=   -  \nabla h + \eps^2R_{s}^{-1} (1+\eps \rho)^{-1} \, \hat{K}\bv, \nonumber \\
&&\rho_{\tau}+  \nabla\cdot\bv +\eps \nabla\cdot (\rho \bv) = 0 . \label{1}
\end{eqnarray}
where $R_s$ in the streaming Reynolds number defined in Section 1.
For small $\eps$, function $h(1+\eps \rho)$ can be written as
\begin{equation}
h=A_{1} \, \rho +\frac{\eps^2}{2}  \, A_{2} \, \rho^{2} +
\frac{\eps^3}{6}  \, A_{3}  \, \rho^{3} + \dots   \label{2}
\end{equation}
where $A_{n}$ ($n=0,1,\dots$) are constants given by
\[
A_{1}=1, \quad
A_{n}=\frac{\rho^{*n}_{0}}{c_0^2} \, \frac{d^{n}h^*(\rho^{*}_{0})}{d\rho^{*n}_{0}}
\quad (n\geq 2).
\]
We are looking for an asymptotic expansion of solutions of Eqs. (\ref{1}) for small $\eps$ and large streaming Reynolds numbers
such that $R_s^{-1}=O(\eps^{\lambda})$ with $\lambda\geq 1$. In this limit, the viscosity does not appear in the first three terms
of the expansion and flows described by those terms are effectively inviscid. Therefore, from now on we will completely
ignore viscous terms.

\section{\label{sec3}Asymptotic expansion}

If we insert (\ref{2}) into (\ref{1}) and then put $\eps=0$, we obtain the standard equations that describe
the linear sound waves in an inviscid compressible fluid. Here we are interested in slow mean motions which coexist with and are affected
by the sound waves.
Therefore, we assume that the mean flow develops
on the slow time scale $t=\eps^2\tau$ and that $\bv$ and $\rho$ are functions of both $\tau$ and $t$.
This choice of the `slow time' is based upon the following two requirements. First, since our aim is to construct
an asymptotic expansion which is valid at least on a time interval of order unity in slow time, it is natural to
maximize this interval by choosing slow time as `slow' as possible, i.e. we maximize $\alpha$ in $t=\eps^{\alpha}\tau$.
Second, the evolution of the mean flow with slow time must be completely determined by asymptotic equations and it can be shown
that if we choose $\alpha>2$, this would lead to asymptotic equations which do not determine
this evolution. So, the only choice satisfying the above requirements is $t=\eps^2\tau$. In Ref. ~\cite{Vladimirov2012} and ~\cite{Vladimirov2010},
this choice of slow time is referred to as a distinguished limit.
It should be emphasized here that the introduction of the slow time is needed only if we want to describe the \emph{unsteady evolution}
of the mean flow.
If the slow time was not introduced we would obtain the steady version of the same averaged equations (Eqs. (\ref{3.46}) below).
So, the introduction of the slow time serves two purposes: (i) to describe the unsteady evolution of the mean flow and (ii) to determine
the natural time scale of this evolution.

With this assumption, Eqs.
(\ref{1}) (without the viscous term) become
\begin{eqnarray}
&&\bv_{\tau}+ \eps^2\bv_{t}+\eps (\bv\cdot\nabla)\bv=   - \nabla h ,  \label{3.1} \\
&&\rho_{\tau}+ \eps^2\rho_{t} + \nabla\cdot\bv +\eps \nabla\cdot (\rho \bv) = 0 .  \label{3.2}
\end{eqnarray}
We seek a solution of these equations in the form of the asymptotic series
\begin{equation}
\bv = \bv_0  +  \eps \bv_1 + \eps^2 \bv_2 +\dots,  \quad
\rho =\rho_0 + \eps \rho_1 + \eps^2 \rho_2 + \dots  \label{3.3}
\end{equation}
Substitution of the second equation (\ref{3.3}) into (\ref{2}) yields
\begin{equation}
h = \rho_{0} +  \eps \left(\rho_{1}+ H_{1}\right)
+  \eps^{2} \left(\rho_{2}+ H_{2}\right) + \dots  \label{3.5}
\end{equation}
where
\begin{equation}
H_{1}= A_2 \, \frac{\rho_0^2}{2}, \quad
H_{2}= A_2 \, \rho_0\rho_1 + A_3 \, \frac{\rho_0^3}{6}, \quad {\rm etc.} \label{3.6}
\end{equation}
Now we substitute (\ref{3.3}) and (\ref{3.5}) into Eqs. (\ref{3.1}) and (\ref{3.2}) and collect the terms of the same power in $\eps$. This results in
the following sequence of equations:
\begin{equation}
\pr_{\tau}\bv_{0}+  \nabla \rho_{0} =0,  \quad
\pr_{\tau}\rho_{0}+ \nabla\cdot\bv_{0}  = 0   \label{3.7}
\end{equation}
and
\begin{equation}
\pr_{\tau}\bv_{k}+  \nabla \rho_{k} =\mathbf{F}_k,  \quad
\pr_{\tau}\rho_{k}+ \nabla\cdot\bv_{k}  = G_k   \label{3.8}
\end{equation}
for $k=1,2,\dots$, where
\[
\mathbf{F}_1 = - \nabla H_1 - (\bv_{0}\cdot\nabla)\bv_{0}, \quad
G_1 = - \nabla \cdot\left(\rho_0\bv_{0}\right),
\]
and where
\begin{eqnarray}
&&\mathbf{F}_k = - \pr_{t}\bv_{k-2} - \nabla H_k - \sum_{l=0}^{k-1}(\bv_{l}\cdot\nabla)\bv_{k-l-1},  \nonumber \\
&&G_k = - \pr_{t}\rho_{k-2}  - \nabla \cdot\left( \sum_{l=0}^{k-1}\rho_l\bv_{k-l-1}\right)  \nonumber
\end{eqnarray}
for $k=2,3,\dots$
Throughout the paper, all functions of $\tau$ are assumed to be $2\pi$-periodic in $\tau$. Therefore, any function
$f(\tau)$ can  be written as
\[
f(\tau)=\overline{f}+\widetilde{f}(\tau), \quad \overline{f}=\frac{1}{2\pi}\int\limits_{0}^{2\pi}f(\tau) d\tau,
\]
where $\overline{f}$ is the mean value of $f(\tau)$ and $\widetilde{f}(\tau)=f(\tau)-\overline{f}$ is the oscillatory part
of $f$.

In what follows, we will be looking for $2\pi$-periodic solutions of equations having the form
(cf. Eqs. (\ref{3.7}), (\ref{3.8}))
\[
\psi_{\tau}(t,\tau)=q(\psi(t,\tau), t,\tau).
\]
Integrating this from $0$ to $2\pi$ in $\tau$ yields the following solvability condition (the necessary condition for existence of
a $2\pi$-periodic solution):
\[
\overline{q(\psi(t,\tau), t,\tau)}=0.
\]
This solvability condition eliminates secular terms in the asymptotic expansion (i.e. terms which grow linearly with $\tau$)
and results in averaged equations.

\subsection{The leading order equations}
Applying averaging to Eqs. (\ref{3.7}), we find that
\begin{equation}
\nabla \overline{\rho}_{0} =0,  \quad
\nabla\cdot\overline{\bv}_{0}  = 0 .  \label{3.11}
\end{equation}
The first of these means that at leading order the averaged density can only depend on the slow time, $\overline{\rho}_0 = \overline{\rho}_0(t)$.
The second equation says that the leading order mean flow must be incompressible. In what follows we choose
\begin{equation}
\overline{\rho}_{0} =0 \label{3.12}
\end{equation}
as the only solution that is physically meaningful.

For oscillatory parts of $\bv_0$ and $\rho_0$, we have
\begin{equation}
\pr_{\tau}\widetilde{\bv}_{0}+  \nabla \widetilde{\rho}_{0} =0,  \quad
\pr_{\tau}\widetilde{\rho}_{0}+ \nabla\cdot\widetilde{\bv}_{0}  = 0 .  \label{3.13}
\end{equation}
Then the first equation (\ref{3.13}) imply that
\begin{equation}
\widetilde{\bv}_{0}=\nabla\phi_0,    \label{3.15}
\end{equation}
while the second equation (\ref{3.13}) leads to the standard wave equation for $\phi_0$:
\begin{equation}
\pr_{\tau}^2\phi_0 - \nabla^2\phi_0=0 .   \label{3.16}
\end{equation}
Thus, the oscillatory part of the leading-order flow is irrotational and represents the usual sound waves.

\subsection{The first-order equations}
On averaging Eqs. (\ref{3.8}) for $k=1$, we find that
\begin{eqnarray}
&&\nabla \overline{\rho}_{1} =  -  \overline{(\bv_0\cdot\nabla)\bv_0} - \nabla \overline{H}_1 ,  \quad \nonumber \\
&&\nabla\cdot\overline{\bv}_{1} = - \nabla\cdot\overline{(\rho_0 \bv_0)} .  \label{3.17}
\end{eqnarray}
Using (\ref{3.15}), the first equation (\ref{3.17}) can be written as
\begin{eqnarray}
(\overline{\bv}_0\cdot\nabla)\overline{\bv}_0 = -\nabla \overline{\Pi}_0, \quad
\overline{\Pi}_0=\overline{\rho}_{1}+ \overline{\vert\nabla\phi_0\vert^2}/2+\overline{H}_1. \label{3.18}
\end{eqnarray}
The first equation (\ref{3.18}) together with the second equation (\ref{3.11}) represent the stationary Euler equations for an inviscid incompressible
fluid, with function $\Pi_0$ playing the role of the pressure. Thus, in general, the leading order averaged flow is described by the
stationary incompressible Euler equations and may coexist with the background sound waves. However, this averaged flow and the sound waves
do not interact. Therefore, we restrict our attention to the case where the leading-order mean flow is absent,
i.e.
\begin{eqnarray}
\overline{\bv}_0=\mathbf{0}, \quad \overline{\Pi}_0=0.  \label{3.20}
\end{eqnarray}
Note that the second equation (\ref{3.20}) implies that
\begin{eqnarray}
\overline{\rho}_{1} = - \overline{\vert\nabla\phi_0\vert^2}/2 - \overline{H}_1 + C_1(t)  \label{3.21}
\end{eqnarray}
where $C_1(t)$ is an arbitrary function which, for each particular problem, can be chosen using boundary conditions.

It follows from the second equation (\ref{3.11}) and from Eqs. (\ref{3.13}) that
\begin{equation}
\nabla\cdot\overline{\bv}_{1} =
\overline{\tilde{\rho}_{0} \, \pr_{\tau}\tilde{\rho}_{0}} + \overline{\tilde{\bv}_{0}\cdot \pr_{\tau}\tilde{\bv}_{0}}=0 . \label{3.22}
\end{equation}
Here the last equality follows from the fact that $\overline{f'(\tau)f(\tau)}=0$ for any $2\pi$-periodic function $f(\tau)$.
Thus, the first-order averaged flow is also incompressible.

Separating the oscillatory part in Eqs. (\ref{3.8}) for $k=1$ and using (\ref{3.20}), we obtain
\begin{eqnarray}
&&\pr_{\tau}\widetilde{\bv}_{1}+  \nabla \widetilde{\rho}_{1} =  -  \widetilde{(\widetilde{\bv}_0\cdot\nabla)\widetilde{\bv}_0}
- \nabla \widetilde{H}_1 ,  \quad \nonumber \\
&&\pr_{\tau}\widetilde{\rho}_{1}+ \nabla\cdot\widetilde{\bv}_{1} = -\nabla\cdot\widetilde{(\widetilde{\rho}_0 \widetilde{\bv}_0)} .  \label{3.23}
\end{eqnarray}
With the help of (\ref{3.13}) and (\ref{3.15}), these can be written as
\begin{eqnarray}
\pr_{\tau}\widetilde{\bv}_{1}+  \nabla \widetilde{\rho}_{1} =  -  \nabla \Phi_1 ,  \quad
\pr_{\tau}\widetilde{\rho}_{1}+ \nabla\cdot\widetilde{\bv}_{1} = \pr_{\tau}\Psi_1  \label{3.24}
\end{eqnarray}
where
\[
\Phi_1 =  \widetilde{\vert\nabla\phi_0\vert^2}/2+\widetilde{H}_1 ,  \quad
\Psi_1 = \widetilde{\vert\nabla\phi_0\vert^2}/2
+\widetilde{(\pr_{\tau}\phi_0)^2}/2 .
\]
Taking $curl$ of the first equation (\ref{3.24}), we find that $\pr_{\tau}(\nabla\times\widetilde{\bv}_1)=\mathbf{0}$, which, in turn,
implies that $\nabla\times\widetilde{\bv}_1=\mathbf{0}$. We conclude that $\widetilde{\bv}_1$ is irrotational, so that
\begin{eqnarray}
\widetilde{\bv}_1 = \nabla\phi_1 . \label{3.26}
\end{eqnarray}

\subsection{The second-order equations}

Averaging the first equation (\ref{3.8}) for $k=2$ and some further manipulations yield
\begin{equation}
\overline{\rho}_{2} = - \overline{\nabla\phi_0\cdot\nabla\phi_1} - \overline{H}_2 + C_2(t) \label{3.27}
\end{equation}
where $C_2(t)$ is an arbitrary function which, for each particular problem, can be chosen with the help of boundary conditions.

The oscillatory part of the first equation (\ref{3.8}) for $k=2$ can be written as
\begin{eqnarray}
&&\pr_{\tau}\widetilde{\bv}_{2}+  \nabla \widetilde{\rho}_{2} = - (\widetilde{\bv}_0\cdot\nabla)\overline{\bv}_1 -
(\overline{\bv}_1\cdot\nabla)\widetilde{\bv}_0 \nonumber \\
&&\qquad\qquad\qquad\quad -  \nabla \left(\pr_t\phi_0 + \nabla\phi_0\cdot\nabla\phi_1+\widetilde{H}_2\right).  \label{3.30}
\end{eqnarray}
Further manipulations yield
\begin{equation}
\pr_{\tau}\widetilde{\bv}_{2}+  \nabla \widetilde{\rho}_{2} = \widetilde{\bv}_0 \times \overline{\bom}_1 -\nabla \Phi_2 \label{3.32}
\end{equation}
where  $\overline{\bom}_1=\nabla\times\overline{\bv}_1$ is the vorticity of the velocity field $\overline{\bv}_1$
and $\Phi_2 = \pr_t\phi_0 + \nabla\phi_0\cdot\nabla\phi_1+\widetilde{H}_2 + \overline{\bv}_1\cdot\widetilde{\bv}_0$.
Taking the $curl$ of (\ref{3.32}) results in the following equation for $\widetilde{\bom}_2=\nabla\times\widetilde{\bv}_2$:
\begin{equation}
\pr_{\tau}\widetilde{\bom}_{2} = \nabla\times\left(\widetilde{\bv}_0 \times \overline{\bom}_1\right). \label{3.34}
\end{equation}

\subsection{The third-order equations}
On averaging the first equation (\ref{3.8}) for $k=3$ and using Eq. (\ref{3.20}), we get
\begin{eqnarray}
\pr_{t}\overline{\bv}_1 + (\overline{\bv}_1\cdot\nabla)\overline{\bv}_1 &=& - \nabla \left(\overline{\rho}_{3}+\overline{H}_3\right) - \overline{(\widetilde{\bv}_0\cdot\nabla)\widetilde{\bv}_2} \nonumber \\
&&- \overline{(\widetilde{\bv}_1\cdot\nabla)\widetilde{\bv}_1} - \overline{(\widetilde{\bv}_2\cdot\nabla)\widetilde{\bv}_0}   .  \label{3.35}
\end{eqnarray}
With the help of (\ref{3.15}) and (\ref{3.26}) this can be simplified to
\begin{eqnarray}
\pr_{t}\overline{\bv}_1 + (\overline{\bv}_1\cdot\nabla)\overline{\bv}_1 = \overline{\widetilde{\bv}_0 \times\widetilde{\bom}_2}
- \nabla \overline{\Pi}^*_1   \label{3.36}
\end{eqnarray}
where
\begin{equation}
\overline{\Pi}_1^* = \overline{\rho}_3+\overline{\widetilde{\bv}_2\cdot\widetilde{\bv}_0} + \overline{\vert\nabla\phi_1\vert^2}/2 + \overline{H}_3. \label{3.37}
\end{equation}
Let $\bxi$ be the field of displacements of fluid particles in the oscillatory velocity field $\widetilde{\bv}_0$ defined by
\begin{equation}
\pr_{\tau}\bxi = \widetilde{\bv}_0, \quad \overline{\bxi}= \mathbf{0}. \label{3.38}
\end{equation}
Then it can be shown that
\begin{eqnarray}
\overline{\widetilde{\bv}_0 \times\widetilde{\bom}_2} &=& - \frac{1}{2}\left(
\overline{\bxi \times \left(\nabla\times\left(\pr_{\tau}\bxi  \times \overline{\bom}_1\right)\right)}\right. \nonumber \\
&&\qquad\quad +\left.\overline{\pr_{\tau}\bxi \times \left(\nabla\times\left( \overline{\bom}_1 \times \bxi \right)\right)}\right). \label{3.40}
\end{eqnarray}
This can be further simplified with the help of the identity
\begin{eqnarray}
\ba\times\left(\nabla\times(\bbb\times\bc)\right) +
\bbb\times\left(\nabla\times(\bc\times\ba)\right) + \, \bc\times\left(\nabla\times(\ba\times\bbb)\right) &=&
\nabla\left(\ba\cdot\left(\bbb\times\bc\right)\right)+(\nabla\cdot\ba) (\bbb\times\bc)  \nonumber \\
&&+(\nabla\cdot\bbb) (\bc\times\ba)+(\nabla\cdot\bc) (\ba\times\bbb), \qquad \label{3.41}
\end{eqnarray}
which is valid for arbitrary vector fields $\ba(\bx)$, $\bbb(\bx)$ and $\bc(\bx)$.
Applying this identity to Eq. (\ref{3.40}), we obtain
\begin{eqnarray}
\overline{\widetilde{\bv}_0 \times\widetilde{\bom}_2} =
\bV\times\overline{\bom}_1
-\frac{1}{2}\nabla\left(\overline{\bom}_1\cdot\overline{\left(\bxi  \times \pr_{\tau}\bxi \right)}\right) \label{3.45}
\end{eqnarray}
where
\begin{equation}
\bV =  \frac{1}{2}\overline{\left[\pr_{\tau}\bxi, \bxi\right]} \label{3.43}
\end{equation}
and, where
$\left[\ba, \bbb\right]  =  (\bbb\cdot\nabla)\ba-(\ba\cdot\nabla)\bbb$
is the commutator of $\ba(\bx)$ and $\bbb(\bx)$.
It can be shown that $\bV$ is the Stokes drift velocity of fluid particles (sometimes also called the Lagrangian drift velocity)
induced by the leading order oscillatory flow
(see, e.g., Eq. (12) in Ref. ~\cite{Westervelt} or Eq. (21) in Ref. ~\cite{Longuet}).
The Stokes drift velocity is obtained
by averaging the velocity of each fluid particle  rather than averaging the Eulerian velocity at a fixed point in space.
The most important feature of the
Stokes drift velocity is that it may be nonzero in a purely oscillatory flow where the mean Eulerian velocity is zero (for more details
see Ref. ~\cite{Longuet}).

Finally, we substitute (\ref{3.45}) into (\ref{3.36}) and obtain the closed system of equations for $\overline{\bv}_1$:
\begin{eqnarray}
\pr_{t}\overline{\bv}_1 + (\overline{\bv}_1\cdot\nabla)\overline{\bv}_1 = \bV\times\overline{\bom}_1  - \nabla \overline{\Pi}_1,
\quad \nabla\cdot\overline{\bv}_1=0 \quad   \label{3.46}
\end{eqnarray}
where
\begin{equation}
\overline{\Pi}_1 = \overline{\Pi}_1^* + \frac{1}{2}\nabla\left(\overline{\bom}_1\cdot\overline{\left(\bxi
\times \pr_{\tau}\bxi \right)}\right). \label{3.47}
\end{equation}
For a given acoustic field,  the Stokes drift velocity can be computed using Eq. (\ref{3.43}). Then Eqs. (\ref{3.46}) can be solved under
appropriate initial and boundary conditions resulting in a mean velocity field $\overline{\bv}_1$ and a function $\overline{\Pi}_1$.
We should emphasize here that we do not need to know what quantities are contained in $\overline{\Pi}_1$, because Eqs. (\ref{3.46}) (supplemented
with appropriate initial and boundary conditions) lead to a function  $\overline{\Pi}_1$ which is unique up to addition of an arbitrary function
of $t$. Once we know $\overline{\Pi}_1$, we can find an expression for $\overline{\rho}_3$ which is similar to Eqs. (\ref{3.21}) and (\ref{3.27}).

The effect of the background sound waves on the mean motion is described by the first term on the right side of the first equation (\ref{3.46}) that
contains the Stokes drift velocity $\bV$. Therefore, we refer to Eqs. (\ref{3.46}) as the \emph{acoustic drift equations}, and these represent
the main result of the paper.

The term $\bV\times\overline{\bom}_1$ that describes the effect of the sound waves on the mean flow may look strange at first sight as it is of fourth order
in the amplitude of the sound waves (both $\bV$ and $\overline{\bom}_1$ are the second-order quantities) but appears in the equation that is quadratic in the amplitude. All possible quadratic interactions had been described more than 50 years ago by Chu and Kov\'{a}sznay \cite{Chu}. In the absence of dissipation, the only quadratic interaction is between two sound modes. However, in agreement with the analysis of Chu and Kov\'{a}sznay \cite{Chu},
it results in a potential force in the averaged equations and does not lead to generation of vorticity. Therefore one needs to go further
and take account of cubic and quartic interactions. The present paper shows that the cubic terms do not affect the averaged flow
and that the only term that
gives a non-trivial contribution is of fourth order in the amplitude of the sound waves. The fact that it arises in the averaged equations
for the (quadratic in the amplitude) mean flow is natural because although the Stokes-drift velocity is small (quadratic in the amplitude), its effect over a long time interval (of order $\eps^{-2}$) is
not small (of order unity), and this is why it appears in Eqs. (\ref{3.46}).

It is interesting to note that equations similar to (\ref{3.46}) had been derived earlier by
Craik and Leibovich to describe the Langmuir circulations in the ocean \cite{Craik}.
In the  case of steady two-dimensional flows, Eqs.
(\ref{3.46}) reduce to the inviscid version of the equations derived by Riley (see Eq. (30) in Ref. ~\cite{Riley2001}) in the context of the steady streaming in an incompressible fluid.

We note also that the Stokes drift velocity $\bV$, given by (\ref{3.43}), is incompressible: $\nabla\cdot\bV=0$. This is a consequence of
the fact that $\tilde{\bv}_{0}$ and $\tilde{\rho}_{0}$ represent acoustic waves and satisfy Eqs. (\ref{3.13}) and (\ref{3.16}).

The above asymptotic expansion allows us to compute successive approximations to the radiation force exerted on a rigid body in an acoustic field
in a lossless medium. An interesting feature of our model is that both the first-order (quadratic in the sound wave amplitude) and the second-order
(cubic) radiation force are not affected by the mean flow despite the fact that the mean velocity is the first-order quantity.
To show this, we consider a fixed rigid body in a lossless medium. The radiation force is the integral of the averaged excess pressure
$\overline{P}^*-P^*_0$ over the body surface. Our procedure yields the following expansion for $\overline{P}^*-P^*_0$:
\begin{eqnarray}
\frac{\overline{P}^*-P^*_0}{\rho^*_0 c^{*2}_0} &=& \eps^2\left(\overline{\rho}_1+
\frac{B_2}{2} \, \overline{\tilde{\rho}_0^2}\right)  \nonumber \\
&&+
\eps^3\left(\overline{\rho}_2+ B_2 \, \overline{\tilde{\rho}_0 \tilde{\rho}_1} + \frac{B_3}{6} \, \overline{\tilde{\rho}_0^3}\right)  \nonumber \\
&&+\eps^4\Bigl(\overline{\rho}_3+ \dots \Bigr) + \dots
\label{3.48}
\end{eqnarray}
Here $B_k$ are constants that can be expressed in terms of $A_k$ defined in Section 2. Note that $\overline{\rho}_1$ and $\overline{\rho}_2$
are given by (\ref{3.21}) and (\ref{3.27}), and $\overline{\rho}_3$ can be found from Eq. (\ref{3.47}) after solving Eq. (\ref{3.46}) with appropriate initial and boundary conditions. It is evident from (\ref{3.21}) and (\ref{3.27}) that neither $\overline{\rho}_1$ nor $\overline{\rho}_2$
depend on the mean velocity $\overline{\bv}_1$, so the first two terms on the right side of (\ref{3.48}) are not affected by the mean flow.
Note also that the first term can be reduced to the standard formula of Ref. ~\cite{WangLee}.

\section{\label{sec4}Example}

To show that sound waves can have a significant effect on slow motions described by Eqs. (\ref{3.46}), we consider
an example which is similar in spirit to Craik's theory of Langmuir circulations \cite{Craik2}. Namely, we will
show that a weak steady flow may become unstable if a simple acoustic field is present.

We assume that the acoustic wave field produces the Stokes drift velocity of the form $\bV=\Phi(x,y)\mathbf{e}_z$ for some function $\Phi$ and consider flows that are independent of $z$, i.e.
\[
\overline{\bv}_1=(u(x,y),v(x,y),w(x,y))\quad \hbox{and} \quad \overline{\Pi}_1=\overline{\Pi}_1(x,y).
\]
In this case, Eqs. (\ref{3.46}) can be written as
\begin{eqnarray}
&&\pr_{t}\bv^{\perp}+(\bv^{\perp}\cdot\nabla^{\perp})\bv^{\perp}=-\nabla^{\perp} \Pi^* - w \nabla^{\perp}\Phi, \nonumber \\
&&\pr_{t}w+(\bv^{\perp}\cdot\nabla^{\perp})w=0, \nonumber \\
&&\nabla^{\perp}\cdot\bv^{\perp}=0, \label{4.1}
\end{eqnarray}
where $\Pi^*=\overline{\Pi}_1- w\Phi$, $\bv^{\perp}=(u,v,0)$ and $\nabla^{\perp}=(\pr_{x},\pr_{y},0)$. Formally, Eqs. (\ref{4.1}) coincide with the equations governing
two-dimensional motion of a stratified fluid in the Boussinesq approximation, with $w$ playing the role of the fluid density and $(-\Phi)$ being the potential of the external body force.

Equations (\ref{4.1}) have steady solutions of the form
\begin{equation}
\bv^{\perp}=0, \quad w=W(x,y), \quad \Pi^*=P(x,y) \label{4.2}
\end{equation}
where $W(x,y)$ and $P(x,y)$ must satisfies
\begin{equation}
W\nabla^{\perp}\Phi=\nabla^{\perp} P. \label{4.3}
\end{equation}
This implies that
\[
\pr_x W \, \pr_y\Phi - \pr_y W \, \pr_x\Phi =0.
\]
The last equality is satisfied if
\begin{equation}
\Phi=F(W) \label{4.4}
\end{equation}
for some function $F$.

Let $\hat{\bv}^{\perp}(x,y,t)$, $\hat{w}(x,y,t)$, $\hat{p}(x,y,t)$ represent
a perturbation of steady state (\ref{4.2}). Assuming that
the perturbation is small, we linearize Eqs. (\ref{4.1}):
\begin{eqnarray}
&&\pr_{t}\hat{\bv}^{\perp}=-\nabla^{\perp} \hat{P} - \hat{w} \, \nabla^{\perp}\Phi, \nonumber \\
&&\pr_{t}\hat{w}+(\hat{\bv}^{\perp}\cdot\nabla^{\perp})W=0, \nonumber \\
&&\nabla^{\perp}\cdot\hat{\bv}^{\perp}=0. \label{4.5}
\end{eqnarray}
The linearized equations conserve the perturbation energy given by
\begin{equation}
E=\frac{1}{2}\int_{\cal D} \left( \left\vert\hat{\bv}^{\perp}\right\vert^2 - \frac{d\Phi}{dW} \, \hat{w}^2\right) dx dy =\mathrm{const}.\label{4.6}
\end{equation}
It is assumed that that the perturbation either sufficiently rapidly decays as $\sqrt{x^2+y^2}\to\infty$ or periodic in $x$ and $y$.
The domain of integration $\mathcal{D}$ is the entire $xy$ plane in the first case and the rectangle of periods in the second case.
If $d\Phi/dW\leq 0$ everywhere in the flow domain, then $E$ is a non-negative quantity and can be used as
a measure of the amplitude of the perturbation,
and the conservation of $E$ implies that the perturbation cannot grow with time, so that steady state (\ref{4.2}) is stable to small
perturbations. The stability corresponds to a `stably stratified equilibrium' in our analogy with the Boussinesq fluid.
If, however, $d\Phi/dW > 0$ in some part of the flow domain, then it can be shown using the technique of Ref. ~\cite{Vladimirov1989}
that perturbations for which $E$ is negative
grow with time $t$ exponentially. In this case, basic state (\ref{4.2}) is unstable, and this instability may lead to nontrivial mean flows
that coexist with the background acoustic field.

To give an explicit example of this instability, we suppose that initially there is a weak steady shear flow
\[
W=\gamma y \, \mathbf{e}_z .
\]
for some $\gamma>0$. (Although this velocity profile is unbounded, it is a good approximation to any shear flow provided
the perturbation is localized in the $y$ direction.) Then we `switch on' an acoustic field in the form of two plane waves propagating at
the angles $\pm\alpha$ to the $z$ axis:
\[
\phi_0=\Re\left(e^{i(\mathbf{k}\cdot\mathbf{x}-\tau)}+ i e^{i(\mathbf{q}\cdot\mathbf{x}-\tau)}\right)
\]
where $\mathbf{k}=(0,\sin\alpha,\cos\alpha)$ and $\mathbf{q}=(0,-\sin\alpha,\cos\alpha)$. Formula (\ref{3.43}) results in the
following expression for the Stokes drift velocity:
\[
\bV=\Phi(y) \, \mathbf{e}_z=\cos\alpha\left[1+\cos 2\alpha \, \sin(2y\sin\alpha)\right] \, \mathbf{e}_z.
\]
Hence,
\[
\frac{d\Phi}{dW}=\frac{\Phi'(y)}{W'(y)}=\frac{\sin 4\alpha}{2\gamma} \cos(2y\sin\alpha).
\]
Evidently, $d\Phi/dW > 0$ for $0< \alpha < \pi/4$ and all $y$ such that $\vert y \vert < \pi/(4\sin\alpha)$. So, if we choose
the initial perturbation such that
$u(x,y,0)=0$, $v(x,y,0)=0$ and $w(x,y,0)$ is nonzero only in the interval $\vert y \vert < \pi/(4\sin\alpha)$, then
the perturbation energy, given by (\ref{4.6}) will be negative, and the perturbation will grow exponentially.
The nonlinear development of this instability will, in turn, lead to nontrivial slow motions coexisting with the fast acoustic field.

\section{Discussion}

By employing the regular asymptotic procedure we have obtained the equations governing the evolution of slow mean
flows that coexist with and are affected by
the background sound waves. It is evident from Eqs. (\ref{3.46}) that if the mean velocity is zero initially, at $t=0$, it will remain zero for
all $t>0$. Therefore,
Eqs. (\ref{3.46}) do not lead to any acoustic streaming if there is no initial perturbation. However, if an initial perturbation is present,
its evolution will be described by Eqs. (\ref{3.46}) and, therefore, will be affected by the sound waves via the drift velocity that appears in
(\ref{3.46}). The example considered in Section IV shows that if initially there is a weak steady shear flow, then a simple acoustic field in the form of two plane
sound waves may result in instability leading to a non-trivial unsteady mean flow. In contrast with the classical theory of acoustic streaming at low streaming
Reynolds numbers, there is no need for attenuation of the sound waves in our theory.

The asymptotic expansion described in the paper is valid if the streaming Reynolds number $R_{s}$ is high.
To be precise, our theory works if $R_s^{-1}=O(\eps^\lambda)$ as $\eps\to 0$ with $\lambda\geq 1$. However, it is not difficult to see that
if $R_s=\mu^{-1}$ for some $\mu=O(1)$, then the effects of viscosity can be incorporated
in our procedure, and the only modification of the averaged equations
 (\ref{3.46}) will be the presence of the viscous term
$\mu \nabla^2\overline{\bv}_1$ on the right side of the first equation (\ref{3.46}).

The acoustic drift equations (\ref{3.46}) are valid only in the regions of the flow domain that
are sufficiently far away from rigid boundaries. If we want to include a rigid boundary, then we have to take into account viscous boundary layers which are essential irrespective of how large the Reynolds number is. For $R_s\lesssim 1$, this can be done following the
approach of Ref. ~\cite{IM2012}.

\textbf{Acknowledgments.} 
{We are grateful to Profs. A. D. D. Craik, S. Leibovich, H. K. Moffatt, and N. Riley
for helpful discussions and to the anonymous referees whose comments helped to improve the original manuscript.}

\end{document}